\documentclass[a4paper,11pt]{article}
\pdfoutput=1 % if your are submitting a pdflatex (i.e. if you have
\usepackage{graphicx}% Include figure files
\usepackage{dcolumn}% Align table columns on decimal point
\usepackage{bm}% bold math
\usepackage{float}
\usepackage{subcaption}
\usepackage{multirow}
\usepackage{comment}
\usepackage{dsfont}
\usepackage{amsthm}
\usepackage{xcolor}
\usepackage{url}
\usepackage{tabularx}
\usepackage{booktabs}
\usepackage{makecell}
\usepackage{jheppub} % for details on the use of the package, please
\newcolumntype{Y}{>{\centering\arraybackslash}X}
\usepackage{pifont}
\newcommand{\cmark}{\ding{51}}%
\newcommand{\xmark}{\ding{55}}%

\newtheorem{proposition}{Proposition}[section]

\DeclareMathOperator{\sign}{sgn}

\usepackage[T1]{fontenc} % if needed

\title{An Efficient Lorentz Equivariant Graph Neural Network for Jet Tagging}% Force line breaks with \\

\author[a,e,1]{Shiqi Gong\note{This work was done when the first author was visiting Microsoft Research Asia.}}
\author[b]{Qi Meng}
\author[b]{Jue Zhang}
\author[c]{Huilin Qu}
\author[d]{Congqiao Li}
\author[d]{Sitian Qian}
\author[a]{Weitao Du}
\author[a]{Zhi-Ming Ma}
\author[b]{Tie-Yan Liu}
\affiliation[a]{Academy of Mathematics and Systems Science, Chinese Academy of Sciences,\\Zhongguancun East Road, Beijing 100190, China}
\affiliation[b]{Microsoft Research Asia,\\Danling Street, Beijing 100080, China}
\affiliation[c]{CERN, EP Department,\\CH-1211 Geneva 23, Switzerland}
\affiliation[d]{School of Physics, Peking University, \\Chengfu Road, Beijing 100871, China}
\affiliation[e]{University of Chinese Academy of Sciences, \\No.19 Yuquan Road, Beijing 100049, China}
\emailAdd{gongshiqi15@mails.ucas.ac.cn}
\emailAdd{meq@microsoft.com}
\emailAdd{tyliu@microsoft.com}

\abstract{Deep learning methods have been increasingly adopted to study jets in particle physics. %Since symmetry-preserving has been shown to be an important factor to boost the performance of deep learning in many applications, Lorentz group equivariance - a fundamental spacetime symmetry for elementary particles has been incorporated in deep learning model for jet tagging.
{Since symmetry-preserving behavior has been shown to be an important
factor for improving the performance of deep learning in many applications,
Lorentz group equivariance - a fundamental spacetime symmetry for elementary
particles - has recently been incorporated into a deep learning model for
jet tagging. However, the design is computationally costly due to the analytic construction
of high-order tensors.}
In this article, we introduce LorentzNet, a new symmetry-preserving deep learning model for jet tagging. The message passing of LorentzNet relies on an efficient Minkowski dot product attention.
Experiments on two representative jet tagging benchmarks show that LorentzNet achieves the best tagging performance and improves significantly over existing state-of-the-art algorithms. 
The preservation of Lorentz symmetry also greatly improves the efficiency and generalization power of the model, allowing LorentzNet to reach highly competitive performance when trained on only a few thousand jets. Code and models are available at \url{https://github.com/sdogsq/LorentzNet-release}.
}

\begin{document} 
\maketitle
\flushbottom

\section{Introduction}
Machine learning techniques, especially deep learning, have been widely adopted to solve a variety of problems in collider physics. Jet tagging, i.e., identifying which type of elementary particle initiates the jet, is one of the most notable applications. It provides a powerful handle for analyzing collision events and serves as a crucial tool for new physics searches and standard model measurements. %(See reference ~\cite{Kogler:2018hem} on a review of the development and use of jet substructure technique). 
Based on the theory and phenomenology of jet physics, many expert-designed high-level jet substructure observables are constructed for jet tagging (See reference ~\cite{Kogler:2018hem,Abdesselam:2010pt,Altheimer:2012mn,Altheimer:2013yza,Adams:2015hiv,Marzani:2019hun,Larkoski:2017jix} for reviews of the development and use of jet substructure technique). 
Using different jet representations, various deep learning approaches have been investigated in recent years: {Linear classifier \cite{Komiske:2017aww} and} multilayer perceptrons (MLPs)~\cite{Datta:2017rhs,Datta:2017lxt,Datta:2019ndh,Chakraborty:2019imr,Chakraborty:2020yfc} can be trained on a collection of jet-level observables; {Fisher discriminant analysis~\cite{Cogan:2014oua}, MLPs~\cite{Almeida:2015jua}, locally connected networks~\cite{Baldi:2016fql}, and 2D convolutional neural networks (CNNs)~\cite{deOliveira:2015xxd,Barnard:2016qma,Komiske:2016rsd,Kasieczka:2017nvn,Macaluso:2018tck,Choi:2018dag,Dreyer:2018nbf,Lin:2018cin,Du:2019civ,Li:2020grn,Filipek:2021qbe,Fraser:2018ieu,CMS:2020poo} are applied to jet images; MLPs~\cite{Guest:2016iqz,Pearkes:2017hku,Butter:2017cot,Kasieczka:2018lwf}, 1D CNNs~\cite{CMS:2020poo}, and recurrent neural networks~\cite{Guest:2016iqz,Egan:2017ojy,Fraser:2018ieu,Bols:2020bkb} are used to process a jet as a sequence of its constituent particles;} Graph neural networks (GNNs)~\cite{Komiske:2018cqr,Qu:2019gqs,Moreno:2019bmu,Moreno:2019neq,Mikuni:2020wpr,Bernreuther:2020vhm,Guo:2020vvt,Dolan:2020qkr,Mikuni:2021pou,Konar:2021zdg,Shimmin:2021pkm} are developed for the ``particle cloud'', i.e., an unordered set of particles; Recursive neural networks and GNNs are adopted to study jet clustering trees~\cite{Louppe:2017ipp,Cheng:2017rdo,Dreyer:2020brq,Dreyer:2021hhr}. Compared with classic jet substructure methods, deep learning approaches can better leverage the fine resolution of modern detectors and automatically discover complex patterns from low-level inputs, leading to substantial improvements in jet tagging performance.

 However, a key aspect neglected by most of these deep learning models is the fundamental spacetime symmetry.
The jet tagging result should not depend on the spatial orientation or the Lorentz boost of a jet. {Starting from the work \cite{cohen2016group}, symmetry-preserving deep learning models have been developed and shown their power in improving performance \cite{zaheer2017deep,finzi2020generalizing,kanwar2020equivariant,boyda2021sampling,satorras2021n,hutchinson2021lietransformer,batzner2021se} and enhancing the model's interpretability \cite{lenc2015understanding} in many applications, e.g., the Deep Sets framework is proposed for preserving the permutation invariance and equivariance \cite{zaheer2017deep}, \cite{kanwar2020equivariant,boyda2021sampling} design gauge equivariant flow for sampling for lattice gauge theory, and \cite{hutchinson2021lietransformer,satorras2021n,batzner2021se} propose group equivariant models for physical or molecular dynamics prediction. For jet tagging, symmetry-preserving deep learning models have not been fully explored.}
%\textcolor{red}{When applying the deep learning models on the tasks where the input to output mapping has symmetry, incorporating the symmetry in the model design has been shown to be important to boost the performance \cite{finzi2020generalizing,kanwar2020equivariant,boyda2021sampling,satorras2021n,hutchinson2021lietransformer,batzner2021se}  and increase the model's interpretability .}
Although a preprocessing step is applied in many existing models to reduce such dependence, e.g., by performing a Lorentz boost along the collision axis and a rotation in the transverse plane such that the jet momentum always points to the same direction, this only functions to a limited extent and cannot achieve full invariance to arbitrary Lorentz transformations. There exist models such as LoLa~\cite{Butter:2017cot} and LBN~\cite{Erdmann:2018shi} that attempt to exploit the underlying Lorentz symmetry with specially designed neural network layers, but they cannot guarantee the invariance of the entire model. A better way to achieve independence from any Lorentz transformation is via equivariant neural networks. A neural network layer is Lorentz-equivariant if its output transforms accordingly when the input undergoes a Lorentz transformation. Therefore, equivariant layers can be stacked to build a symmetry-preserving deep neural network, of which the classification result is unchanged under any Lorentz transformation.
In \cite{bogatskiy2020lorentz}, LGN, a Lorentz equivariant neural network architecture built with tensor products of Lorentz group representations is %proposed
{developed and demonstrated for the first time}. {However, the tagging performance of LGN reported in \cite{bogatskiy2020lorentz} is {unsatisfactory}. One bottleneck is the slow computational speed due to the explicit calculation of the high-order tensors through consecutive tensor product operations over hidden layers. } 

In this article, we propose a new design of symmetry-preserving deep learning models for jet tagging. Using the particle cloud representation of a jet, we propose LorentzNet, which directly scalarizes the input 4-vectors to realize Lorentz symmetry. Specifically, we design Minkowski dot product attention, which aggregates the 4-vectors with weights learned from Lorentz-invariant geometric quantities under the Minkowski metric. 
% The scalars are fed into the deep neural networks to ensure the underlying weights can be well approximated. 
The construction of LorentzNet is guided by the universal approximation theory on Lorentz equivariant mapping, which ensures the equivariance and the universality of the LorentzNet. {Compared to LGN which requires computationally expensive tensor products of the {geometric quantities} to achieve expressiveness, LorentzNet only requires the Minkowski inner product of two vectors.}  Thus, it is more efficient in terms of both training and inference.
Benchmarked on two representative jet tagging datasets, LorentzNet achieves the best tagging performance and improves significantly over existing state-of-the-art algorithms such as ParticleNet. %Moreover, even with 0.5\% of training samples, LorentzNet still achieves competitive performance on the top tagging dataset, which clearly shows the benefit of the inductive bias brought by the Lorentz symmetry. 
{Moreover, we test the tagging performance with 5\%, 1\% and 0.5\% of training samples, the degradation in performance for LorentzNet is smaller than that for ParticleNet,} which clearly shows the benefit of the inductive bias brought by the Lorentz symmetry.

The rest of this paper is organized as follows. Sec.~\ref{MethodSec} reviews the theory of the Lorentz group. Sec.~\ref{NetworkArch} introduces the architecture of LorentzNet. Sec.~\ref{Experiment} details the experiments carried out on two jet tagging benchmarks to show the effectiveness of LorentzNet. Sec.~\ref{Conclusion} serves as the conclusion.

\section{Preliminary}
\label{MethodSec}
%\subsection{Background}\label{sec:background}
In this section, we introduce the foundation of the Lorentz group and describe the graph neural network for modelling particle clouds. 
\subsection{Notations}
Following the jet representation in \cite{Qu:2019gqs}, we regard the constituent particles as a point cloud, which is an unordered, permutation invariant set of particles.
Let $V=(v_1,\cdots,v_N)\in\mathbb{R}^{N\times 4}$ denote the point clouds living in $\mathbb{R}^4$ and $N$ denotes the number of particles in a jet. Note that the number of particles for jets may be different.  $\langle\cdot,\cdot\rangle$ and $\|\cdot\|$ denote the Minkowski inner product and Minkowski norm. $\oplus$ denotes the direct sum and $\otimes$ denotes the tensor product. $[N]$ denotes the set $\{1,2,\cdots, N\}$. 

\subsection{Theory of the Lorentz Group}
\paragraph{Minkowski metric.} Consider the 4-dimensional space-time $\mathbb{R}^4$ with basis $\{e_i\}_{i=0}^3$. We define a bilinear form $\eta:\mathbb{R}^4\times \mathbb{R}^4\rightarrow \mathbb{R}$ as follows. For $u,v\in\mathbb{R}^4$, we set $\eta(u,v)=u^TJv$ where 
$J=diag(1,-1,-1,-1)$ is the Minkowski metric. The Minkowski inner product of two vectors $u = (t,x,y,z)$ and $v = (t',x',y',z')$ is defined as $\langle u, v \rangle = \eta(u,v) =tt'-xx'-yy'-zz'$.
The Minkowski norm of a vector $u = (t,x,y,z)$ is defined to be $\|u\| = \sqrt{\eta(u,u)} = \sqrt{t^2-x^2-y^2-z^2}$.%$\|v\|=\sqrt{|t^2-x^2-y^2-z^2| }$. 

\paragraph{Lorentz transformation.}
The Lorentz group is defined to be the set of all matrices that preserve the bilinear form $\eta$. Restricting the inertial frames to be positively oriented and positively time-oriented, we obtain proper orthochronous Lorentz group, denoted as $SO(1,3)^+$. The 3d spatial rotation group $SO(3)$ is a subgroup of $SO(1,3)^+$. Additionally, the Lorentz boost is also included. Given two inertial frames $\{e_i\}_{i=0}^3$ and $\{e_i'\}_{i=0}^3$, the relative velocity vector $\beta$ and the boost factor $\gamma$ are defined by $e'_0=\gamma e_0+\sum_{i=1}^3\gamma\beta_ie_i$ where $\gamma=(1-\beta^2)^{-1/2}$. 

If we perform Lorentz boost along the x-spatial axis, then the Lorentz transformation between these two frames is the matrix 
$$Q=\left(\begin{matrix}\gamma&-\gamma\beta&0&0\\-\gamma\beta&\gamma&0&0\\0&0&1&0\\0&0&0&1 \end{matrix}\right).$$

\paragraph{Lorentz group equivariance.}
Let $T_g: V\rightarrow V$ and $S_g: U\rightarrow U$ be group actions of $g \in G$ on sets $V$ and $U$, respectively. We say a function $\phi: V\rightarrow U$ is equivariant to group $G$ if
\begin{equation}
    \phi(T_g(v))=S_g(\phi(v))
\end{equation}
holds for all  $v \in V$ and $g \in G$.

In this work, we only consider the case that the type of the output is a scalar or vector. Therefore, we explore the following equivariance on a set of particles $V\in\mathbb{R}^{N\times 4}$. Let $Q$ be the Lorentz transformation, the Lorentz equivariance of $\phi(\cdot)$ means: 
\begin{align}
    Q\phi(v)=&\phi(Qv), \quad \textit{for $\phi(v)\in \mathbb{R}^4$;}\\
    \phi(v)=&\phi(Qv), \quad  \textit{for $\phi(v)\in\mathbb{R}$.}
\end{align}Note that when the output is a scalar, the group equivariance equals the group invariance.

\subsection{Graph Neural Network for Particle Cloud}
A jet can be denoted as a graph when we regard the constituent particles as nodes.
For the particle with index $i$, we use its 4-momentum vector $v_i=(E^i,p_x^i,p_y^i,p_z^i)$ as the coordinate of node $i$ in Minkowski space. We use $s_i=(s^i_1,s^i_2,\cdots,s^i_\alpha)$ to denote the scalars, such as mass, charge and particle identity information, etc, which compose the node attributes. 
Now $f_i = v_i\oplus s_i$ contains essential features for tagging.
%For each particle, we use its 4-momentum vector $v=(E,p_x,p_y,p_z)$ as the coordinate of a node in Minkowski space. \textcolor{blue}{We use $s=(s_1,s_2,\cdots,s_\alpha)$ to denote the scalars such as mass, charge and particle identity information, etc, which compose the node attributes. We index the particles in a jet as $f_i$ and $f_i$ contains essential features for tagging, which is denoted as $f_i=v_i\oplus s_i$. 
%where $x_i=(E^i, p_x^i,p_y^i,p_z^i)$ denotes the 4-momentum vector and $s_i=(s_1^i,s_2^i,\cdots,s_\alpha^i)$ denotes the sets of scalars. 
%}
The graph can be denoted as $G=(V,E)$ where $V$ is the set of nodes and $E$ is the set of edges. The edges characterize the message passing between two particles, hence the interaction of two individual sets of particle-wise features. If there is no such interaction, there will be no edge between the two corresponding nodes. Here, we regard the graph as a fully connected graph as we do not assume that we have any prior on the interactions among these particles.

Graph neural networks are natural to learn representations for graph-structured data  \cite{scarselli2008graph}. Given a graph $G=(V,E)$, assuming $L$ steps in total, the $l$-th message passing step on the graph {can be described as} \cite{gilmer2017neural}:
% Given a graph $G=(V,E)$, 1
\begin{align}
   & m_i^{l+1}=\sum_{j\in \mathcal{N}(i)} M_l(h_i^l,h_j^l,e_{ij});\\
    & h_i^{l+1}=U_l(h_i^l,m_i^{l+1});
\end{align}
%\textcolor{blue}{where $l$ denotes the message passing step, 
where $h_i^0=f_i$ is the input feature, $e_{ij}$ is the edge feature, $\mathcal{N}(i)$ is the set of neighbors of the node $i$, and $M_l, U_l$ are neural networks.
%}
For a classification problem, the output $\hat{y}$ can be obtained by applying the softmax function after decoding $\{h_i^L;i\in [N]\}$.

Directly applying the message passing to the jet can not ensure the Lorentz group equivariance because the non-linear neural networks are applied directly to the whole input features and ignore its intrinsic structure. 
There are some variants of message passing designed to satisfy continuous group symmetry. A common way is to project the inputs to the basis of irreducible representations of the corresponding group, e.g., the LGN \cite{bogatskiy2020lorentz} which satisfies the Lorentz group equivariance. To ensure the universality, {high-order geometric tensors} should be realized by LGN, which brings high computational cost. (See detailed discussions in Appendix \ref{App:LGN}). 
%Radial Field \cite{kohler2019equivariant} and EGNN \cite{satorras2021n} for E(n)-equivariance, TFN \cite{thomas2018tensor} for SE(3) equivariance and Schnet \cite{schutt2018schnet} for E(n)-invariance. However, no one of them can directly satisfy the Lorentz group equivariance. %\textcolor{red}{The LGN \cite{bogatskiy2020lorentz} is proposed to satisfy Lorentz group symmetry, which projects the inputs to the bases of irreducible representations of Lorentz group via tensor product. }

In the next section, we will introduce the architecture of LorentzNet which is built upon a universal approximation theorem of the Lorentz equivariant function and does not require explicitly calculating the higher-order representations.
%which adopts the message passing framework in EGNN to ensure the Lorentz group equivariance. 

\section{Network Architecture}
\label{NetworkArch}
In this section, we illustrate the architecture of LorentzNet. %We follow the framework of EGNN \cite{satorras2021n} and change the layers and norm to satisfy Lorentz group equivariance and better adapt to jet tagging tasks. 
The construction of the LorentzNet is based on the following universal approximation theorem for the Lorentz group equivariant continuous function. 

\begin{proposition}\label{prop}\cite{villar2021scalars} 
A continuous function $\phi:(\mathbb{R}^{N\times 4})\rightarrow \mathbb{R}^4$ is Lorentz-equivariant if and only if 
\begin{align}
    \phi(v_1,v_2,\cdots,v_N)=\sum_{i=1}^N g_i(\langle v_i,v_j \rangle_{i,j=1}^N)v_i,\label{eq:5}
\end{align}where $g_i$ are continuous Lorentz-invariant scalar functions, and $\langle\cdot,\cdot\rangle$ is the Minkowski inner product. 
\end{proposition}
Proposition~\ref{prop} provides a way to construct Lorentz group equivariant mapping with no need to calculate the high-order tensors. Instead, a Lorentz group equivariant continuous mapping can be constructed by the attention on $v_i$ with encoding the Minkowski dot products of $v_i$ with its neighbours. This motivates us to design the Minkowski dot product attention in LorentzNet, which will be introduced in the next section.

%In the following sections, we will construct the Lorentz equivariant mapping based on Proposition \ref{prop}.
\begin{figure*}
    \centering
    \includegraphics[width=1\linewidth]{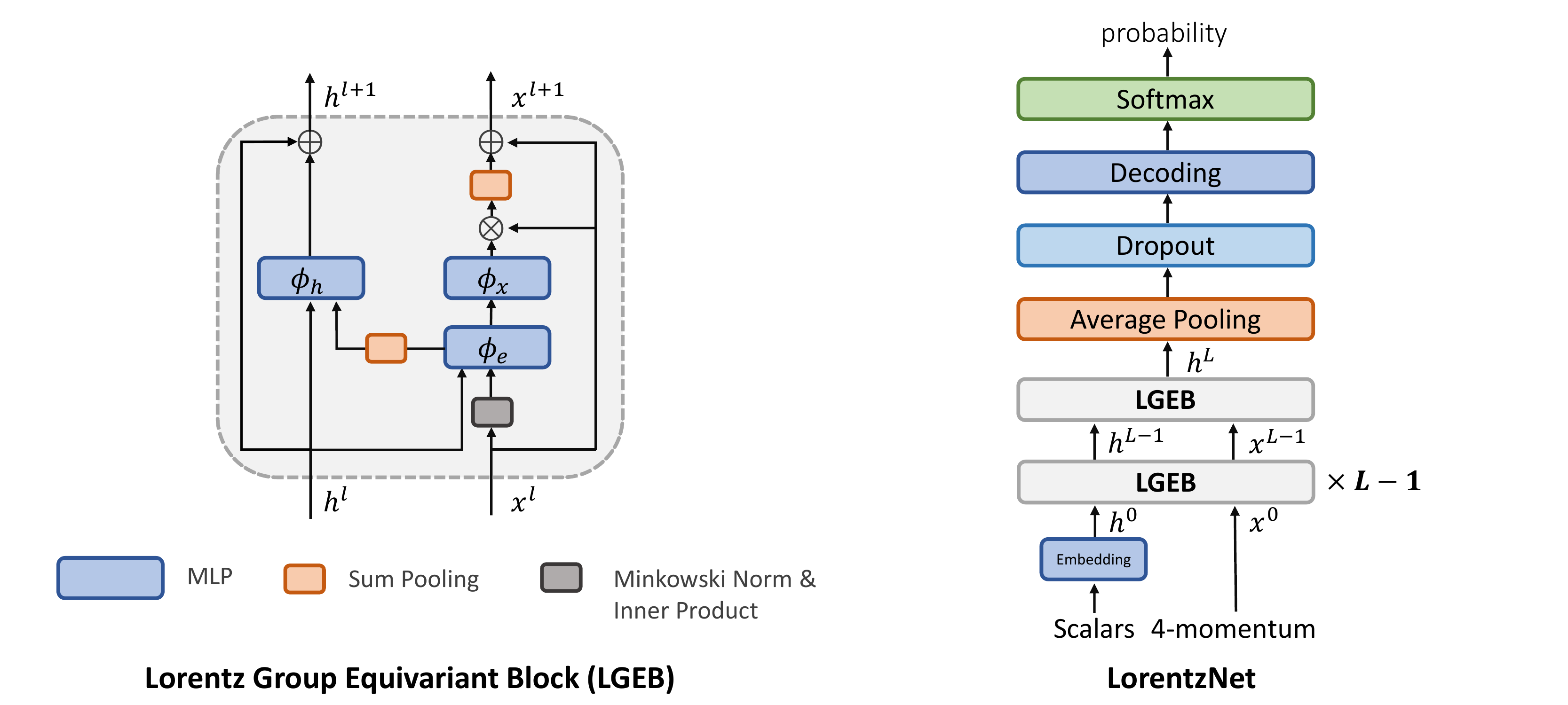}
    \caption{(\textbf{left}): The structure of the Lorentz Group Equivariant Block (LGEB). (\textbf{right}): The network architecture of the LorentzNet. }
    \label{fig:lorentznet}
\end{figure*}
\subsection{LorentzNet}
We introduce the blocks in LorentzNet. As described in Fig.~\ref{fig:lorentznet}, LorentzNet is mainly constructed by the stack of Lorentz Group Equivariant block (LGEB) along with encoder and decoder layers. 

\paragraph{Input layer.}
The inputs into the network are 4-momenta of particles from a collision event, and may include scalars associated with them (such as label, charge, etc.). Using the notations defined in Section 2.3, the input is the set of vectors $f_i=v_i\oplus s_i$. %where $x_i=(E^i, p_x^i,p_y^i,p_z^i)$ denotes the 4-momentum vector and $s_i=(s_1^i,s_2^i,\cdots,s_\alpha^i)$ is the collection of scalars.
In the experiments in this paper, the scalars include the mass of the particle (i.e., $(E^i)^2-(p_x^i)^2-(p_y^i)^2-(p_z^i)^2$) %and a label distinguishing observed decay products from the collider beams, or 
or particle identification (PID) information directly if it is available.

\paragraph{Lorentz Group Equivariant Block.}
We use $h^l = (h^l_1,h^l_2,\cdots,h^l_N)$ to denote the node embedding scalars, and $x^l = (x^l_1,x^l_2,\cdots,x^l_N)$ to denote the coordinate embedding vectors in the $l$-th LGEB layer. When $l=0$, $x_i^0$ equals the input of the 4-momenta $v_i$ and $h_i^0$ equals the embedded input of the scalar variables $s_i$. LGEB aims to learn deeper embeddings $h^{l+1}, x^{l+1}$ via current $h^{l}, x^{l}$.  Motivated by  Equation~(\ref{eq:5}), the message passing of LorentzNet is written as follows. We use $m_{ij}$ to denote the edge message between particle $i$ and $j$, and it encodes the scalar information of the particle $i$ and $j$, i.e.,
\begin{align}
    & m_{ij}^l=\phi_e\left(h_i^l,h_j^l,\psi(\|x_i^l-x_j^l\|^2), \psi(\langle x_i^l,x_j^l\rangle)\right), \label{eq:mij}
\end{align}where {$\phi_e(\cdot)$ is a neural network and $\psi(\cdot)=\sign(\cdot) \log(|\cdot| + 1)$ in Equation~(\ref{eq:mij}) is to {normalize large numbers from broad distributions for ease of optimization.}} %to make the heavy tailed distributed quantities centralized for ease of optimization.
Except for the embedding of the scalar features $h_i^l$ and $h_j^l$, according to Proposition~\ref{prop}, the input of the neural network contains the Minkowski dot product $\langle x_i,x_j\rangle$. The $\|x_i^l-x_j^l\|^2$ is also included because the interaction between particles relies on this term and we include it as a prior feature for ease of learning.

According to Equation~(\ref{eq:5}), we design \emph{Minkowski dot product attention} as
\begin{align}
    & x_i^{l+1}=x_i^l+c\sum_{j \in [N]}\phi_x(m^l_{ij})\cdot x_j^l \label{eq:x}
\end{align}where %$\mathcal{N}(i)$ is the neighbors of particle $i$
 %$\psi$ is a fixed scaling function and 
$\phi_x(\cdot)\in\mathbb{R}$ is a scalar function modeled by neural networks. To ensure the equivariance, we can not arbitrarily apply the normalization trick to control the scale of $x_i^{l+1}$. Therefore, %we introduce the hyperparameter $c$ to control the forward stability together with the shortcut connection. 
{the hyperparameter $c$ is introduced to control the scale of $x_i^{l+1}$ to avoid the scale exploding.} This step captures the interactions of the $i$-th particle with other particles via the ensemble of the 4-momenta of all particles. Unlike most of the symmetry-preserving neural networks such as LGN and EGNN \cite{satorras2021n} (for E(n) equivariance)\footnote{The relation with EGNN is discussed in the Appendix~\ref{app:egnn}.} which only {apply non-linear transformation (e.g., neural network)} to encode the radial distance $\|x_i-x_i\|^2$, %as the only scalars extracted from the vector representation, 
we include the dot product $\langle x_i,x_j\rangle$ in $m_{ij}$ according to Equation~(\ref{eq:5}). %which can not be captured by the radial distance. 

The scalar features for particle $i$ is forward as
\begin{align}
    %& m_i^l=\sum_{j\neq i}w_{ij}m^l_{ij} \label{eq:mi}\\
    &h_i^{l+1}= h_i^{l} + \phi_h(h_i^l,\sum_{j\in [N]}w_{ij}m^l_{ij}), \label{eq:h}
\end{align}where $\phi_h(\cdot)$ is also modeled by neural networks whose output dimension equals the dimension of $h_i^{l+1}$. For efficient computation, we operate summation $\sum_{j \in [N]}w_{ij}m_{ij}^l$ to aggregate $m_{ij}^l$. {We introduce an neural network $\phi_m(\cdot)$ to learn the edge significance from node $j$ to node $i$, i.e., $w_{ij}=\phi_m(m^l_{ij}) \in [0,1]$.} This can both ensure the permutation invariance and also ease the implementation for jets with different numbers of particles. This operation is also widely adopted in other types of graph neural networks \cite{gilmer2017neural, satorras2021n}.
%where %$\mathcal{N}(i)$ is the neighbors of particle $i$
%$c$ is a hyperparameter to control the forward stability, %$\psi$ is a fixed scaling function and 
%$\phi_e, \phi_x, \phi_h$ are modeled by neural networks.
%In Equation~(\ref{eq:mij}), we use $m_{ij}$ to denote the edge message between particle $i$ and $j$, and we use the norm $\|\cdot\|^2$ in Minkowski space. %In Equation~(\ref{eq:mi}), we use $m_i$ to denote the node message by aggregating the edge messages along with other nodes. 
%In Equation~(\ref{eq:x}) and (\ref{eq:h}), we update the node and coordinate embeddings by messages and shortcut connection. 

\paragraph{Decoding layer.}
After stacks of LGEB for $L$ layers, we decode the node embedding $h^L=(h_1^L,\cdots,h_N^L)$. Note that the information of $x^{L-1}$ has been included in $h^L$ through the $m_{ij}^{L-1}$. Therefore, to avoid redundant information, we only decode $h^L$.  First we use average pooling to get
\begin{align}
    h^{av}=\frac{1}{N} \sum_{i\in [N]}h_i^L.
\end{align}A subsequent dropout layer is applied to $h^{av}$ to prevent overfitting. A decoding block with two fully connected layers, followed by a softmax function, is used to generate the output for the binary classification task.

%\textcolor{red}{Shiqi, could you please help to provide more details? e.g., the layer of the three networks, the activation, the dropout, the decoding layer, the softmax...}

\subsection{Theoretical Analysis}\label{sec:B}
In this section, we analyze the Lorentz group equivariance of LorentzNet.

\begin{proposition}
The coordinate embedding $x^l=(x_1^l,x_2^l,\cdots,x_N^l)$ are Lorentz group equivariant and the node embedding $h^{l}=(h_1^l,\cdots,h_N^l)$ are Lorentz group invariant.
\end{proposition}
\textbf{Proof:} We denote $Q$ as the  Lorentz transformation.
If $m_{ij}^l$ are invariant under $Q$ for all $i,j,l$, $x_i^{l+1}$ will be Lorentz group equivariant because
\begin{align*}
    Qx_i^{l+1}&=Q(x_i^l+c\sum_{j \in [N]}x_j^l \cdot \phi_x(m_{ij}))\\
    &=Qx_i^l+c\sum_{j\in [N]}Qx_j^l \cdot \phi_x(m_{ij})).
\end{align*}
Then we illustrate the invariance of $m_{ij}^l$. Let's start from the input. Since Minkowski norm and Minkowski inner product are invariant to Lorentz group, we have $\|x_i^0-x_j^0\|^2=\|Qx_i^0-Qx_j^0\|^2$ and $\langle x_i^0, x_j^0 \rangle = \langle Qx_i^0, Qx_j^0 \rangle$.% Since the 4-momentum vector is Lorentz group equivariant, we have $\|x_i^0-x_j^0\|^2=\|Qx_i^0-Qx_j^0\|^2$ because the determinant of $Q$ equals 1, and similar for $\langle x_i^0, x_j^0\rangle$. 
Therefore, the input of $\phi_e$ are invariant variables under transformation $Q$ and then $m_{ij}^0$ are invariant. Recursively using the invariance of $m_{ij}^l$ and the equivariance of $x_i^l$, we can get the conclusion. $\Box$

As for the expressiveness of the LGEB structure, we have the following discussions. %First, according to the update rule that %According to Proposition \ref{prop}, $\phi_x(\cdot)$ should be an expressive function with input $\{\langle x_i,x_j\rangle_{i,j=1}^N\}$. The universal approximation theorem of neural networks guarantee the expressiveness power and we next expand the update rule in Eq.(\ref{eq:x}) to check whether the input information of LorentzNet contains all the information of $\{\langle x_i,x_j\rangle_{i,j=1}^N\}$.
Because $h_i$ is a function of the aggregation of $m_{ij}$, it contains the information of all the Minkowski dot products of the particle pairs.  Therefore, the weights in Minkowski dot product attention $\phi_x(\cdot)$ in Equation~\eqref{eq:x} at $l$-th layer ($l>1$) contain all the Minkowski dot products of the particle pairs. According to Proposition \ref{prop}, the input information of the attention factor is complete for expressiveness.  
%\begin{align}
%    x_i^{l+1}=x_i^l+c\sum_{j\neq i}\phi_x(\phi_e(h_i^l,h_j^l,\psi(\|x_i^l-x_j^l\|^2),\psi(\langle x_i^l,x_j^l\rangle))),
%\end{align}where $h_i^l=h_i^{l-1}+\phi_h(h_i^{l-1},\sum_{j\neq i}w_{ij}m_{ij})$. Therefore, $x_i^{l+1}$ contains information $\{m_{ij}; i,j\in[N]\}$. Because $m_{ij}$ is a function of the Minkowski inner product $\langle x_i,x_j\rangle$, $x_i^{l+1}$ contains all the pairwised Minkowski inner product of the 4-momentum vectors. 

\subsection{Implementation Details} \label{sec:implement}
The LorentzNet architecture used in this paper is shown in Fig.~\ref{fig:lorentznet}. It consists of 6 Lorentz group equivariant blocks ($L=6$). The scalar embedding is implemented as one fully connected layer which maps the scalars to latent space of width 72, i.e., \textsc{Linear}$(\text{scalar\_num},72)$. $\phi_x, \phi_e, \phi_h$ are all implemented as \textsc{Linear}$(72,72)$ $\rightarrow$ \textsc{ReLU} $\rightarrow$ \textsc{BatchNorm1d}$(72)$ $\rightarrow$ \textsc{Linear}$(72,72)$. {$\phi_m$ is  implemented as \textsc{Linear}$(72,1)$ $\rightarrow$ \textsc{Sigmoid}.} The decoding layers are \textsc{Linear}$(72,72)$ $\rightarrow$ \textsc{ReLU} $\rightarrow$ \textsc{Linear}$(72,2)$. The dropout rate here is 0.2. The hyperparameter $c$ in Equation~(\ref{eq:x}) is chosen to be $5\times 10^{-3}$ and $1\times 10^{-3}$ for the top tagging and the quark-gluon tagging task respectively to achieve the best performances.

The networks are implemented with PyTorch and the training is performed on a cluster with four Nvidia Tesla V100 GPUs.  A batch size of 32 is used on every single GPU for LorentzNet architecture. The \textsc{AdamW} \cite{loshchilov2018decoupled} optimizer, with a weight decay of 0.01, is used to minimize the cross-entropy loss. The LorentzNet is trained for 35 epochs in total. Firstly a linear warm-up period of 4 epochs is applied to reach the initial learning rate $1\times 10^{-3}$. Then a \textsc{CosineAnnealingWarmRestarts} \cite{loshchilov2016sgdr} learning rate schedule with $T_0=4,\,T_{mult}=2$ is adopted for next $28$ epochs. Finally, an exponential learning rate decay with $\gamma=0.5$ is used for the last 3 epochs. We test the model on the validation dataset at the end of each training epoch, and the model with the highest validation accuracy is saved as our best model for the final test. Code and models are available at \url{https://github.com/sdogsq/LorentzNet-release}.

\section{Experiment}
\label{Experiment}
The performance of the LorentzNet architecture is evaluated on two representative jet tagging tasks: top tagging and quark-gluon tagging. In this section, we show the benchmark results.
\subsection{Datasets}
\paragraph{Top tagging.}
We first evaluate LorentzNet on top tagging classification experiments with the publicly available reference dataset \cite{kasieczka2019top}. This dataset contains 1.2M training entries, 400k validation entries, and 400k testing entries. Each of these entries represents a single jet whose origin is either an energetic top quark, a light quark, or a gluon. The events are produced using the \texttt{PYTHIA8} Monte Carlo event generator. The ATLAS detector response is modelled with the \texttt{DELPHES} software package. 

The jets in the reference dataset are clustered using the anti-$k_\mathrm{T}$ algorithm, with a radius of $R = 0.8$. For each jet, the 4-momenta are saved in Cartesian coordinates $(E,p_x,p_y,p_z)$ for up to 200 constituent particles selected by the highest transverse momentum. Following the settings in \cite{bogatskiy2020lorentz}, the colliding particle beams aligned along the z-axis are added. Each jet contains an average of 50 particles, and events with less than 200 are zero-padded.

\paragraph{Quark-gluon tagging.}
The second dataset is for quark-gluon tagging, i.e., discriminating jets initiated by quarks and by gluons. The signal (quark) and background (gluon) jets are generated with \texttt{PYTHIA8} using the $Z(\rightarrow \nu\nu)+(u,d,s)$ and $Z(\rightarrow\nu\nu)+g$ processes, respectively. No detector simulation is performed. The final state non-neutrino particles are clustered into jets using the anti-$k_T$ algorithm with $R=0.4$. This dataset consists of 2 million jets in total, with half signal and half background. We follow the setting in \cite{Qu:2019gqs} to split 1.6M/200K/200K for training, validation, and testing.

One difference of the quark-gluon tagging dataset is that it includes not only the 4-momentum vector but also the type of each particle (i.e., electron, photon, pion, etc.). {We use the one-hot encoding of the type of each particle as input scalars into the LorentzNet.} We include this task to test the performance of LorentzNet under this different type of input feature. %Different from the non-equivariant models such as ParticleNet which uses log-scaled features, LorentzNet can not scale the tensors freely in order to keep the Lorentz equivariance.

\subsection{Baselines and Tagging Performance}

We compare the performance of LorentzNet with six baseline models: ResNeXt-50 \cite{xie2017aggregated}, P-CNN \cite{cms2017boosted}, PFN \cite{Komiske:2018cqr}, ParticleNet \cite{Qu:2019gqs}, EGNN \cite{satorras2021n} and LGN \cite{bogatskiy2020lorentz}. The tagging accuracy of these six models except EGNN has been reported in \cite{Qu:2019gqs} and \cite{bogatskiy2020lorentz}. In this paper, we also investigate their robustness under Lorentz transformation and the computational cost. For self-contained, we briefly introduce them here. The ResNeXt-50 model is a 50-layer convolutional neural network with skip connections for image classification. Representing jets as images, we can apply ResNeXt-50 to jet tagging. The P-CNN is a 14-layer 1D CNN using particle sequences as inputs. The P-CNN architecture is proposed in the CMS particle-based DNN boosted jet tagger and shows significant improvement in performance compared to a traditional tagger using boosted decision trees and jet-level observables. 
The Particle Flow Network (PFN) and the ParticleNet also treat a jet as an unordered set of particles. The PFN is based on the Deep Sets framework. The ParticleNet is based on Dynamic Graph Convolutional Neural Network with carefully designed EdgeConv operation. We also include EGNN as a representative symmetry-preserving model as our baseline which is $E(4)$ equivariant. It is different from LorentzNet in that EGNN uses metrics in Euclidean space instead of Minkowski space. We compare LorentzNet with EGNN to show the necessity of Lorentz group symmetry. For a fair comparison, we set the number of parameters of EGNN in the same order as LorentzNet.

For ResNeXt, P-CNN, PFN, and ParticleNet, we follow the implementation in \cite{Qu:2019gqs}. For LGN, we follow the implementation in \cite{bogatskiy2020lorentz} on top tagging. For the quark-gluon dataset, the implementation details are reported in Appendix \ref{App:LGN}. %\footnote{The batch size we use here is larger than that is reported in the original paper for accelerating the training process.}

% \begin{table*}[t]
% \centering
% \resizebox{\columnwidth}{!}{\begin{tabular}{|c|c|c|c|c|c|c|}
% \hline
%   Model & Equivariance           & Accuracy & AUC      & $1/\varepsilon_B \,\, (\varepsilon_S=0.5)$ & $1/\varepsilon_B \,\, (\varepsilon_S=0.3)$  & \#Params\\ \hline
% ResNeXt     & \xmark & $0.936$  & $0.9837$ & $302 \pm 5$                           & $1147 \pm 58$                    &   1.46M  \\
% P-CNN       & \xmark & $0.930$  & $0.9803$ & $201 \pm 4$                           & $759 \pm 24$                     &   348k   \\
% PFN         & \xmark & $0.932$      & $0.9819$ & $247 \pm 3$                           & $888 \pm 17$                     &   82k  \\
% ParticleNet & \xmark & $0.940$  & $0.9858$ & $397\pm 7$                            & $1615 \pm 93$                    &   366k   \\
% EGNN         & E(4) & $0.922$  & $0.9760$ & $148 \pm 8$                                   & $540 \pm 49$                     &   222k  \\
% LGN         & SO$^{+}$(1,3) & $0.929$  & $0.9640$ & $124 \pm 20$                                   & $435 \pm 95$                     &   4.5k  \\ \hline
% LorentzNet  & SO$^{+}$(1,3) & $\bm{0.942}$  & $\bm{0.9868}$ & $\bm{498 \pm 18}$          & $\bm{2195 \pm 173}$               &   224k  \\                     \hline
% \end{tabular}}
\begin{table*}[htp]
\centering
\begin{tabular}{|c|c|c|c|c|}
\hline
  Model    & Accuracy & AUC  & \makecell{$1/\varepsilon_B$ \\$(\varepsilon_S=0.5)$} & \makecell{$1/\varepsilon_B$ \\$(\varepsilon_S=0.3)$}  \\ \hline
ResNeXt      & $0.936$  & $0.9837$ & $302 \pm 5$                           & $1147 \pm 58$                    \\
P-CNN        & $0.930$  & $0.9803$ & $201 \pm 4$                           & $759 \pm 24$                      \\
PFN          & $0.932$      & $0.9819$ & $247 \pm 3$                           & $888 \pm 17$                    \\
ParticleNet  & $0.940$  & $0.9858$ & $397\pm 7$                            & $1615 \pm 93$                    \\
EGNN        & $0.922$  & $0.9760$ & $148 \pm 8$                                   & $540 \pm 49$               \\
LGN         & $0.929$  & $0.9640$ & $124 \pm 20$                                   & $435 \pm 95$                 \\ \hline
LorentzNet  & $\bm{0.942}$  & $\bm{0.9868}$ & $\bm{498 \pm 18}$          & $\bm{2195 \pm 173}$                \\                     \hline
\end{tabular}
\caption{Performance comparison between LorentzNet and other representative algorithms on top tagging dataset. The results for LorentzNet and EGNN are averaged on 6 runs. The results for other baselines are referred to \cite{Qu:2019gqs,bogatskiy2020lorentz,Komiske:2018cqr}.  }\label{tab1}
\end{table*}

\begin{table*}[htp]
\centering

% \begin{tabular}{|c|c|c|c|c|}
% \hline
%      Model       & Accuracy & AUC    & \makecell{$1/\varepsilon_B$ \\$(\varepsilon_S=0.5)$} & \makecell{$1/\varepsilon_B$ \\$(\varepsilon_S=0.3)$} \\ \hline
% ResNeXt       & $0.811$  & $0.8859$ & $26.9$                                & $71.3$                                \\
% P-CNN       & $0.826$  & $0.9000$ & $35.1$                                & $92.4$                                \\
% PFN         & $0.822$      & $0.8974$ & $32.6$                        & $85.1$                                   \\
% ParticleNet & $0.838$  & $0.9102$ & $39.5$                        & $98.8$                        \\ 
% {EGNN} & $0.803$  & $0.8806$ & $26.3 \pm 0.3$ & $76.6 \pm 0.5$  \\
% LGN & $0.803$  & $0.8141$ & $8.30$ & $15.2$  \\ \hline
% LorentzNet  & $\bm{0.844}$  & $\bm{0.9156}$ & $\bm{42.4 \pm 0.4}$     & $\bm{110.2 \pm 1.3}$  \\                     \hline            
% \end{tabular}

\begin{tabular}{|c|c|c|c|c|}
\hline
     Model       & Accuracy & AUC    & \makecell{$1/\varepsilon_B$ \\$(\varepsilon_S=0.5)$} & \makecell{$1/\varepsilon_B$ \\$(\varepsilon_S=0.3)$} \\ \hline
ResNeXt       & $0.821$  & $0.8960$ & $30.9$                                & $80.8$                                \\
P-CNN       & $0.827$  & $0.9002$ & $34.7$                                & $91.0$                                \\
PFN         & $-$      & $0.9005$ & $34.7 \pm 0.4$                        & $-$                                   \\
ParticleNet & $0.840$  & $0.9116$ & $39.8 \pm 0.2$                        & $98.6 \pm 1.3$                        \\ 
{EGNN} & $0.803$  & $0.8806$ & $26.3 \pm 0.3$ & $76.6 \pm 0.5$  \\
LGN & $0.803$  & $0.8324$ & $16.0$ & $44.3$  \\ \hline
LorentzNet  & $\bm{0.844}$  & $\bm{0.9156}$ & $\bm{42.4 \pm 0.4}$     & $\bm{110.2 \pm 1.3}$  \\                     \hline            
\end{tabular}
\caption{Performance comparison between LorentzNet and other representative algorithms on quark-gluon tagging dataset. The results for LorentzNet, EGNN and LGN are averaged on 6 runs. The results for other baselines are referred to \cite{Qu:2019gqs,Komiske:2018cqr}. }%The results for other baselines are reproduced according to the settings in \cite{qu2020jet,komiske2019energy}.}
\label{tab2}
\end{table*}

\begin{figure}[hbp]
    \centering
    \includegraphics[width=0.45\linewidth]{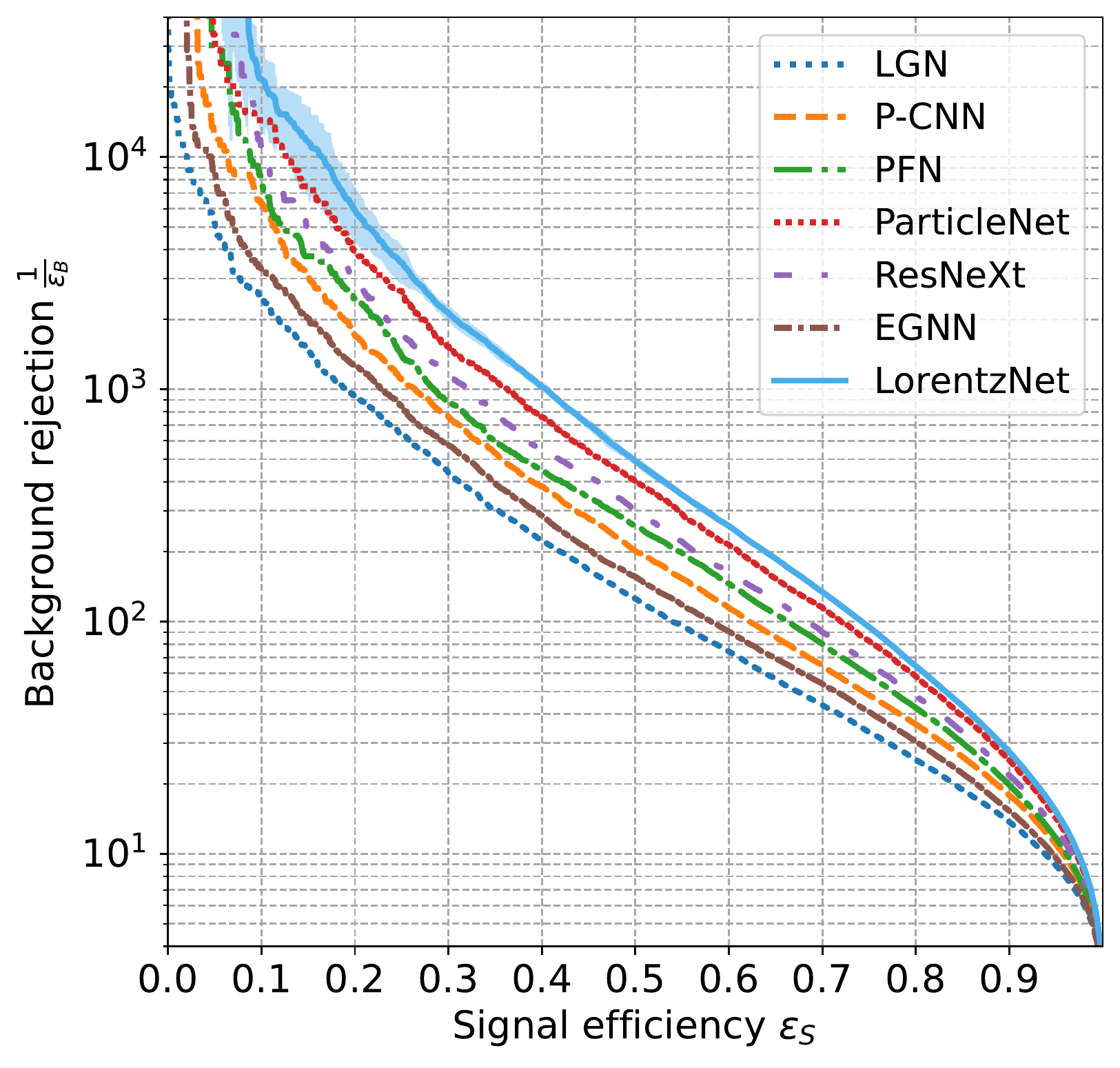}
    \includegraphics[width=0.45\linewidth]{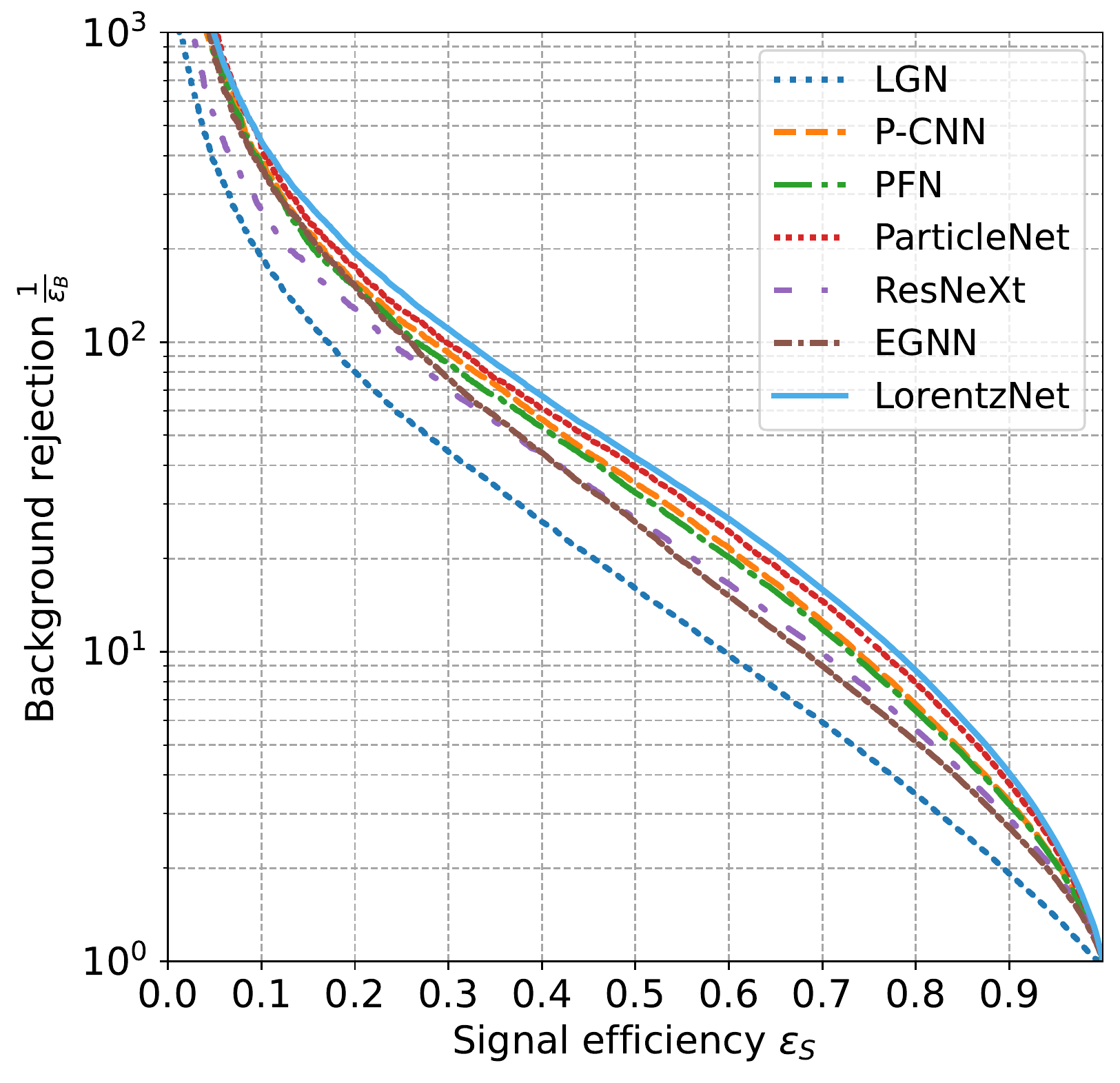}
    \caption{A comparison of ROC curves between LorentzNet and other algorithms on top tagging dataset (left) and quark-gluon dataset (right).}
    \label{fig:my_label}
\end{figure}

\paragraph{Tagging performance.} The results for the top tagging dataset and quark-gluon dataset are summarized in Table \ref{tab1} and Table \ref{tab2}, respectively. We adopt the widely used measures to evaluate the performance of the LorentzNet model including accuracy, the Area Under the ROC Curve (AUC)\footnote{ROC curve is a plot of the true positive rate (Sensitivity) in the function of the false positive rate (100-Specificity) for different cut-off points of a parameter. Each point on the ROC curve represents a sensitivity/specificity pair corresponding to a particular decision threshold. The Area Under the ROC curve (AUC) is a measure of how well a parameter can distinguish between two diagnostic groups.}, the background rejection $1/\varepsilon_B$ at the signal efficiency of $\varepsilon_S=0.3$ and $0.5$ ($\varepsilon_B,\,\varepsilon_S$ are also known as the false positive and the true positive rates, respectively), and the number of trainable parameters.  Especially,
the background rejection metric is widely adopted to select the best jet tagging algorithm as it is directly related to the expected contribution of the background \cite{Qu:2019gqs,bogatskiy2020lorentz,Komiske:2018cqr}.

From Table \ref{tab1} and Table \ref{tab2}, we conclude that LorentzNet achieves the state-of-the-art performance on both the top tagging dataset and the quark-gluon in terms of accuracy, AUC, and background rejection at $\varepsilon_S=0.3,\,0.5$. The results verify the effectiveness of LorentzNet compared with the baselines. 
Fig.~\ref{fig:my_label} shows the background rejection at a fine-grained signal efficiency. The ROC curves of LorentzNet achieve the highest score at all the selected signal efficiency compared to the baselines. {Especially, LorentzNet shows superiority compared to LGN, since it achieves 4 or 5 times improvement on the background rejection. The results verify our discussions in Sec.~\ref{sec:B}.}

\begin{table*}[htp]
% \begin{tabularx}{\linewidth}{YYYYYY}
% \toprule
% Training Fraction & Model & Accuracy & AUC      & $1/\varepsilon_B \,\, (\varepsilon_S=0.5)$ & $1/\varepsilon_B \,\, (\varepsilon_S=0.3)$ \\ \midrule
% \multirow{2}{*}{$0.5\%$} & ParticleNet       & $0.913$  & $0.9687$ & $77 \pm 4$                                & $199 \pm 14$         \\
%  & {LorentzNet} & $\bm{0.929}$  & $\bm{0.9793}$ & $\bm{176 \pm 14}$ & $\bm{562 \pm 72}$  \\ \midrule
% \multirow{2}{*}{$1\%$} & ParticleNet       & $0.919$  & $0.9734$ & $103 \pm 5$                                & $287 \pm 19$         \\
%  & {LorentzNet} & $\bm{0.932}$  & $\bm{0.9812}$ & $\bm{209 \pm 5}$ & $\bm{697 \pm 58}$  \\ \midrule
% \multirow{2}{*}{$5\%$} & ParticleNet & $0.931$  & $0.9807$ & $195\pm 4$ & $609\pm 35$  \\ 
%  & LorentzNet  & $\bm{0.937}$  & $\bm{0.9839}$ & $\bm{293 \pm 12}$     & $\bm{1108 \pm 84}$  \\                     \bottomrule            
% \end{tabularx}
\centering
\begin{tabular}{|c|c|c|c|c|c|}
\hline
\makecell{Training\\Fraction} & Model & Accuracy & AUC      & \makecell{$1/\varepsilon_B$ \\$(\varepsilon_S=0.5)$} & \makecell{$1/\varepsilon_B$ \\$(\varepsilon_S=0.3)$}  \\ \hline
\multirow{2}{*}{$0.5\%$} & ParticleNet       & $0.913$  & $0.9687$ & $77 \pm 4$                                & $199 \pm 14$         \\
 & {LorentzNet} & $\bm{0.929}$  & $\bm{0.9793}$ & $\bm{176 \pm 14}$ & $\bm{562 \pm 72}$  \\ \hline
\multirow{2}{*}{$1\%$} & ParticleNet       & $0.919$  & $0.9734$ & $103 \pm 5$                                & $287 \pm 19$         \\
 & {LorentzNet} & $\bm{0.932}$  & $\bm{0.9812}$ & $\bm{209 \pm 5}$ & $\bm{697 \pm 58}$  \\ \hline
\multirow{2}{*}{$5\%$} & ParticleNet & $0.931$  & $0.9807$ & $195\pm 4$ & $609\pm 35$  \\ 
 & LorentzNet  & $\bm{0.937}$  & $\bm{0.9839}$ & $\bm{293 \pm 12}$     & $\bm{1108 \pm 84}$  \\                     \hline            
\end{tabular}
\caption{Performance comparison between LorentzNet and ParticleNet on top tagging dataset by a fraction of training data. The results are all averaged on 6 runs.}\label{tab12}
\end{table*}

\subsection{Sample Efficiency}
The benefit of the preservation of Lorentz group symmetry in jet tagging has not been studied in the literature. In theory, the Lorentz group symmetry injects inductive bias into the deep learning model which restricts the function class of the hypothesis space. The inductive bias can help to boost the generalization and improve the sample efficiency. As the improvement in the generalization performance (i.e., the tagging accuracy) has been shown in the previous section, we show the robustness of LorentzNet trained on smaller training data to verify the sample efficiency of LorentzNet in this part. 

We choose the best-performed architecture among the models with and without fully Lorentz group symmetry (i.e., the LorentzNet and the ParticleNet) to compare. The inductive bias in ParticleNet is a subgroup symmetry of the Lorentz group, which only considers the Lorentz boosts on the $z$-axis and the rotation on the $x-y$ plane, while LorentzNet is symmetric to the Lorentz group. We randomly select $5\%$, $1\%$, and $0.5\%$ fraction of training data to train the LorentzNet and ParticleNet on the top tagging dataset, and we test their performances on the same test data with size 400k. The training strategy keeps the same as the experiments on the full training data. The results are reported in Table~\ref{tab12}. The gap between the tagging accuracy and AUC between LorentzNet and ParticleNet becomes larger as the number of training data becomes smaller. The results clearly show the benefit of the preservation of Lorentz group symmetry in jet tagging. %indicates the robustness of LorentzNet trained on a smaller dataset.

\subsection{Equivariance test}
\begin{figure}[h]
    \centering
    \includegraphics[width=0.5\linewidth]{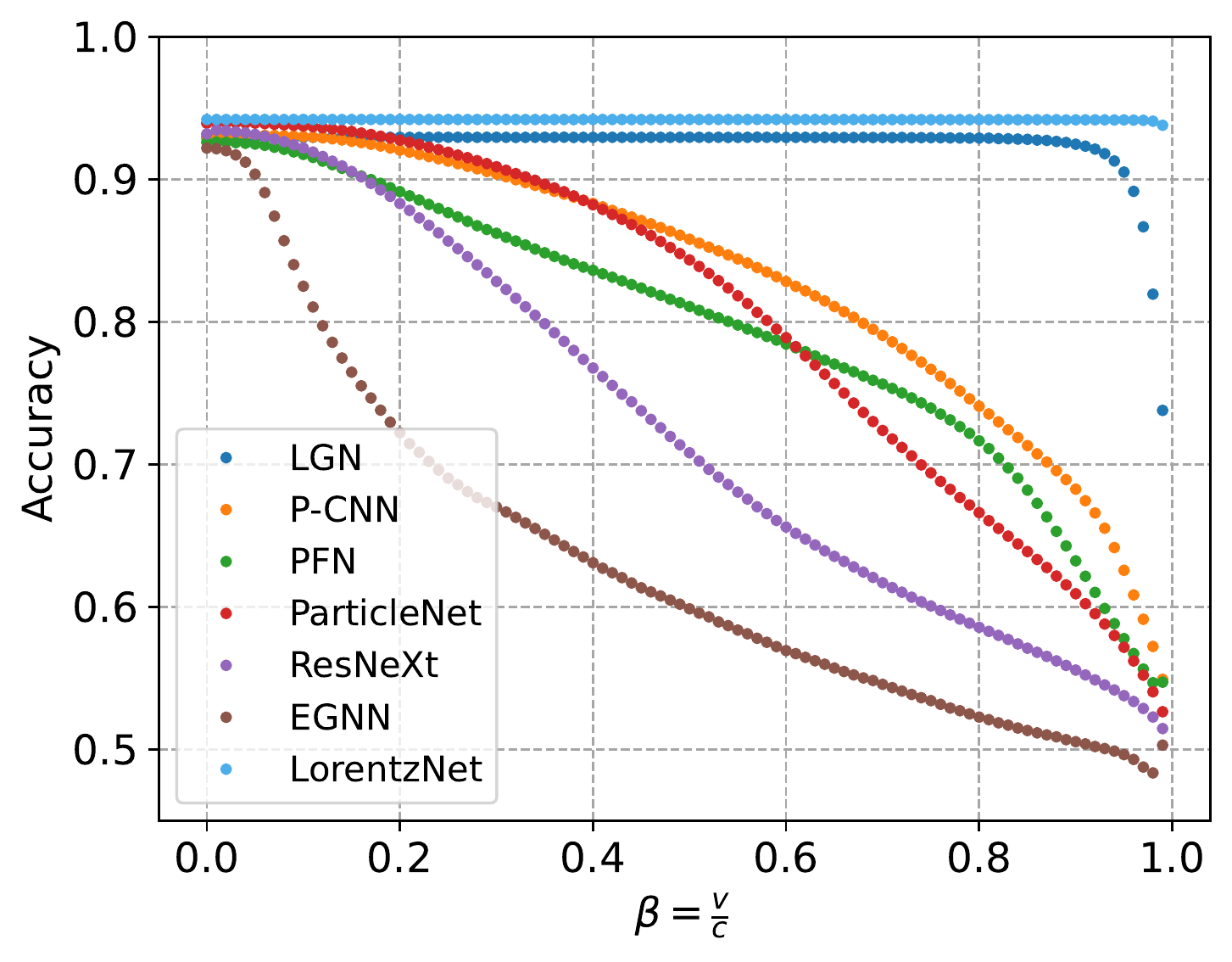}
    \caption{Equivariant test under Lorentz boosts on top tagging dataset.}
    \label{fig:2}
\end{figure}

Another advantage of symmetry-preserving deep learning models is their robustness under Lorentz transformation. To verify it, we rotate the test data by Lorentz transformation with different scales of $\beta$ along the $x-$axis, i.e., the value of $(E, p_x)$ in the 4-momentum vector will be rotated. As $\beta$ becomes larger, the difference between the distributions of training and test data will become larger. We test the model trained on the original training data, and the tagging accuracy on the rotated test data is reported in Fig.~\ref{fig:2}. The horizontal axis of Fig.~\ref{fig:2} shows the value of $\beta$ and the vertical axis shows the tagging accuracy on the top tagging dataset under Lorentz transformation with corresponding $\beta$.
The results show that the accuracy of LorentzNet and LGN on the test data after Lorentz transformation is robust in a large range of $\beta$, while the test accuracy of other non-equivariant models will drop as $\beta$ becomes larger. According to special relativity, the fundamental quantities to clarify the particles will not be changed. %The results show that only the Lorentz group equivariant models LorentzNet and LGN can capture this symmetry. 
Even compared with LGN, LorentzNet is more stable when $\beta$ approaches $1$, and the instability of LGN is caused by the rounding errors in float arithmetic as described in its original paper \cite{bogatskiy2020lorentz}.

%\begin{figure}[h]
%    \centering
%    \includegraphics[width=\linewidth]{figure/QG_ROC_egnn.pdf}
%    \caption{A comparison of ROC curves between LorentzNet and other algorithms on quark-gluon tagging dataset.}
%    \label{fig:my_label2}
%\end{figure}

\begin{table*}[t]
\centering
% \begin{tabularx}{\linewidth}{YcYcYYc}
% \toprule
%             & Equivariance           & Accuracy & AUC      & $1/\varepsilon_B \,\, (\varepsilon_S=0.5)$ & $1/\varepsilon_B \,\, (\varepsilon_S=0.3)$  \\ \midrule
% LorentzNet (w/o)    & \xmark & $0.934$  & $0.9832$ & $290 \pm 30$                           & $1105 \pm 59$                      \\
% LorentzNet  & \cmark & $\bm{0.942}$  & $\bm{0.9868}$ & $\bm{498 \pm 18}$          & $\bm{2195 \pm 173}$                \\                     \bottomrule
% \end{tabularx}
\begin{tabular}{|l|c|c|c|c|c|}
\hline
\multicolumn{1}{|c|}{Model} & Equivariance           & Accuracy & AUC      & \makecell{$1/\varepsilon_B$ \\$(\varepsilon_S=0.5)$} & \makecell{$1/\varepsilon_B$ \\$(\varepsilon_S=0.3)$}  \\ \hline
LorentzNet (w/o)    & \xmark & $0.934$  & $0.9832$ & $290 \pm 30$                           & $1105 \pm 59$                      \\
LorentzNet  & \cmark & $\bm{0.942}$  & $\bm{0.9868}$ & $\bm{498 \pm 18}$          & $\bm{2195 \pm 173}$                \\                     \hline
\end{tabular}
\caption{Performance comparison between LorentzNet and corresponding non-equivalent version on top tagging dataset. Both of the results are averaged on 6 runs.} \label{table:ablation}
\end{table*}

\subsection{Ablation study}
In this section, we report the results of the ablation study to further demonstrate the effectiveness of the components in LorentzNet.
To show the effectiveness of keeping the Lorentz group equivariance, we {directly use} $x_i^l,x_j^l$ as inputs of the $\phi_e$ to break the Lorentz group equivariance because $x_i^l, x_j^l$ are not Lorentz group invariant variables, i.e., the Equation (\ref{eq:mij}) is replaced by
{
\begin{align}
    m_{ij}=\phi_e(x_i^l,x_j^l,h_i^l,h_j^l),%,\psi(\|x_i^l-x_j^l\|^2),\psi(\langle x_i^l,x_j^l\rangle)).
\end{align}}We name this variant as LorentzNet without equivariance (abbreviated as LorentzNet (w/o)).
We compare the performance of LorentzNet (w/o) with LorentzNet on the top tagging dataset. The hyperparameters of training LorentzNet (w/o) keep the same as LorentzNet and we train LorentzNet (w/o) till it converges. As shown in Table~\ref{table:ablation}, LorentzNet (w/o) performs worse than LorentzNet, which shows the necessity of the Lorentz group equivariance on the tagging performance. 

\begin{table*}[t]
\centering
% \begin{tabularx}{\linewidth}{YcYcYYc}
% \toprule
%             & Equivariance  & Time on CPU (ms/batch) & Time on GPU (ms/batch) & \#Params\\ \midrule
% ResNeXt     & \xmark & $5.5$  & $0.34$  &   1.46M  \\
% P-CNN       & \xmark & $\bm{0.6}$  & $\bm{0.11}$  &   348k   \\
% PFN         & \xmark & $\bm{0.6}$  & $0.12$  &   82k  \\
% ParticleNet & \xmark & $11.0$  & $0.19$ &   366k   \\
% {EGNN}         & E(4) & $30.0$  & $0.30$ &   222k  \\
% LGN         & SO$^{+}$(1,3) & $51.4$  & $1.66$ &   4.5k  \\ \midrule
% LorentzNet  & SO$^{+}$(1,3) & $32.9$  & $0.34$ &   224k  \\                     \bottomrule
% \end{tabularx}
\begin{tabular}{|c|c|c|c|c|c|c|}
\hline
Model & Equivariance  & \makecell{Time on CPU \\ (ms/batch)} & \makecell{Time on GPU \\ (ms/batch)} & \#Params\\ \hline
ResNeXt     & \xmark & $5.5$  & $0.34$  &   1.46M  \\
P-CNN       & \xmark & $\bm{0.6}$  & $\bm{0.11}$  &   348k   \\
PFN         & \xmark & $\bm{0.6}$  & $0.12$  &   82k  \\
ParticleNet & \xmark & $11.0$  & $0.19$ &   366k   \\
{EGNN}         & E(4) & $30.0$  & $0.30$ &   222k  \\
LGN         & SO$^{+}$(1,3) & $51.4$  & $1.66$ &   4.5k  \\ \hline
LorentzNet  & SO$^{+}$(1,3) & $32.9$  & $0.34$ &   224k  \\                     \hline
\end{tabular}
\caption{Inference time of each model on both CPU and GPU along with their parameter numbers. Models are executed on a cluster with an Intel Xeon CPU E5-2698 v4 and an Nvidia Tesla V100 32GB. All the inference times are collected with a batch size of 100.}\label{tab3}
\end{table*}
\subsection{Computational cost}
We report the inference time of LorentzNet and other baseline models. The inference time of each model on both CPU and GPU along with the number of their parameters are reported in Table \ref{tab3}. The number of trainable parameters of LorentzNet is in the same order as P-CNN and ParticleNet. Models are executed on a cluster with an Intel Xeon CPU E5-2698 v4 and an Nvidia Tesla V100 32GB GPU. All the inference times are collected with a batch size of 100.  As shown in Table \ref{tab3}, the inference time on GPU of LorentzNet is slightly larger than P-CNN, PFN, and particleNet, and is comparable with ResNeXt. Especially, compared with LGN, the LorentzNet is almost $5$ times faster on GPU, although the number of trainable parameters of LGN is much smaller. For the inference time on CPU, LorentzNet is also faster than LGN. Both results demonstrate the efficiency of LorentzNet. Since there is no need to compute the high-order tensors in LorentzNet, it is more efficient. {See Appendix \ref{App:Cost} for more computational comparisons of LorentzNet and LGN.} The main computational cost of LorentzNet comes from its message passing on the fully connected graph. Its cost quadratically depends on the number of particles in a jet. This cost can be reduced by clustering the nodes and we will explore it in future studies.

\section{Conclusion}
\label{Conclusion}
%We have presented the LorentzNet, a Lorentz group equivariant graph neural network for jet tagging. We showed that LorentzNet can achieve the best accuracy over all existing methods. {Although we evaluate the effectiveness of LorentzNet for the jet tagging problem on two datasets, the whole model architecture may be applied to other tasks on jets, such as, regression - predicting the true jet mass/momentum; generation - simulating jet data, as well as the full collision event, classifying signal vs background processes, pileup mitigation, etc.}  
%As ever-larger science and engineering datasets become available for particle physics, we hope that LorentzNet will help enable more accurate prediction as well as aid the discovery of interesting new phenomena. 
We have presented LorentzNet, a Lorentz group equivariant graph neural network for jet tagging. Experiments on two representative jet tagging datasets show that LorentzNet achieves substantial performance improvement over all existing methods. {The efficient design of the message passing to preserve the Lorentz symmetry significantly reduces both training and inference time compared to LGN. Moreover, the efficient design} enhances the generalization power, allowing LorentzNet to reach highly competitive performance with only a few thousand jets for training. Although the effectiveness of LorentzNet is only demonstrated on two jet tagging tasks in this article, the symmetry-preserving architecture can be applied to a broad range of tasks in particle physics, such as regression of the true jet mass or even the full 4-momentum, generation of jets for fast simulation, classification of the entire collision event, mitigation of pileup interactions, and more.
As ever-larger datasets become available in particle physics, we hope that LorentzNet will help enable more accurate prediction and accelerate the discovery of interesting new phenomena.

\acknowledgments
We thank Prof. Qing-Hong Cao (Peking University) for his helpful discussions on AI methods in physical system modeling.
We also thank Wei Chen (MSRA) for her helpful discussions on symmetry-preserving deep learning models and support.

\bibliographystyle{jhepref}
\bibliography{ref.bib}

\appendix

\section{Details about LGN}\label{App:LGN}
In this section, we first introduce the architecture of LGN and analyze its stability and computational efficiency. 

The input of LGN is the same as LorentzNet. The node features $h^{l+1}$ for LGN is obtained by
\begin{equation}\label{eq:LGN}
    h_i^l\oplus CG\left(h_i^l\otimes h_i^l\right)\oplus CG\left(\sum_j \phi_x(\|x_i-x_j\|^2)(x_i-x_j)\otimes h_j^{l}\right),
\end{equation}where $CG(\cdot\otimes\cdot)$ denotes the  Clebsch-Gordan tensor product and $\phi_x(\cdot)$ denotes a neural network. The term $h_i^l$ is the shortcut connection. The middle term reflects the self-interaction and the last term reflects the interaction of the $i$-th particle with other particles. A multi-layer perceptron layer will be added to operate on the scalar representations of the Lorentz group after the operating Equation~(\ref{eq:LGN}).  The Clebsch-Gordan tensor product projects tensors to different orders of an irreducible representation of the Lorentz group, i.e.,
\begin{align}
 e_{l,m}^{(k,n)}\rightarrow \sum H_{(k,n),l,m}^{(k_1,n_1),l_1,m_1; (k_2,n_2),l_2,m_2}e_{l_1,m_1}^{(k_1,n_1)}\otimes e_{l_2,m_2}^{(k_2,n_2)},\label{eq:A2}
\end{align}where $H_{(k,n),l,m}^{(k_1,n_1),l_1,m_1; (k_2,n_2),l_2,m_2}$ denotes the Clebsch-Gordan coefficient, $e_{(l,m)}^{(k,n)}$ denotes the canonical basis of the $(k,n)$-order irreducible representation of the Lorentz group where $l=|k-n|/2,\cdots,(k+n)/2; m=-l,\cdots, l$. 

Ideally, the universal approximation of the LGN can be achieved if the order of tensors that the LGN realized is sufficiently high. However, in practice, the order of tensors should be cut off for efficient computation and there is no direct prior to determining this cutoff in many applications.

In terms of computational efficiency, one problem is that the tensor product brings the computational cost to (Eq.(\ref{eq:A2})) compared with the simple dot product in LorentzNet.  
Another problem is that the tensor product operation is a weighted sum of products of the input features. Without normalization, the output of the network will be sensitive to the scale of the input, which leads to instability (e.g., value explosion or vanishing) in both the forward and backward processes. 
To ensure stability, the input 4-momenta were scaled by a factor of $0.005$ on the top tagging dataset in the original paper \cite{bogatskiy2020lorentz}.  The scaling factor is a hyper-parameter to tune for different datasets. However, there is no principle way designed for the scaling or normalization which can stabilize the training process and keep the Lorentz symmetry meanwhile.  
Therefore, compared with LGN, LorentzNet can achieve a better tradeoff between approximation ability and computational efficiency.

In the experiments of LGN on the quark-gluon dataset, we follow its public code base in \texttt{https://github.com/fizisist/LorentzGroupNetwork}. The batch size is set to be 48 and we run $53$ epochs in total. Other parameters are all aligned with its original paper \cite{bogatskiy2020lorentz}. The wall block time of running one epoch of LGN is more than $10$ times larger than other baseline models, which is also reported in its original paper \cite{bogatskiy2020lorentz} (e.g., 7.5 hours per epoch with a batch size of $8$ on top tagging). The computational efficiency is the main bottleneck for further evaluation of LGN.

\section{Relation with EGNN} \label{app:egnn}
The message passing of LorentzNet is also inspired by EGNN, which is proposed to ensure $E(n)$ group equivariance. If the output is a scalar, the message passing of EGNN is written as:
\begin{align}
    m_{ij}^{l+1}&=\phi_e(h_i^l,h_j^l,\|x_i-x_j\|_E);\\
    h_i^{l+1}&=\phi_h(h_i^l,\sum_{j\neq i}m_{ij}^l),
\end{align}where $\|\cdot\|_E$ denotes the $L^2$ norm under Euclidean metric.

We discuss the difference between LorentzNet and EGNN. First, EGNN uses the Euclidean norm of the relative distance as the only scalar information of the vector. Since relative distance can not recover the information of the angles (i.e., the inner products), its expressiveness is arguable \cite{du2021equivariant}. In the context of $E(n)$ equivariant, the inner product $\langle x_i,x_j\rangle$ can not be added because of the translation invariance. As we target Lorentz group symmetry, we can include the Minkowski inner product for information completeness. Second, EGNN does not include the coordinate update component for scalar prediction, while LorentzNet includes the Minkowski dot product attention for expressiveness. For fair comparison in the experiments, we include the coordinate update step $x_i^{l+1}=x_i^l+\sum_{j\neq i}(x_j-x_i)\phi_x(m_{ij}^l)$ in EGNN to ensure the difference only caused by different groups.

In the experiments of EGNN on top and quark-gluon dataset, we follow its public code base in \texttt{https://github.com/vgsatorras/egnn}. We use a batch size of 128 in the training process for 35 epochs in total. The \textsc{AdamW} \cite{loshchilov2018decoupled} optimizer, with a weight decay of 0.01, is used to minimize the cross-entropy loss. Firstly a linear warm-up period of 4 epochs is applied to reach the initial learning rate $1\times 10^{-3}$. Then a \textsc{CosineAnnealingWarmRestarts} \cite{loshchilov2016sgdr} learning rate schedule with $T_0=4,\,T_{mult}=2$ is adopted for next $28$ epochs. Finally, an exponential learning rate decay with $\gamma=0.5$ is used for the last 3 epochs. We test the model on the validation dataset at the end of each training epoch, and the model with the highest validation accuracy is saved as our best model for the final test.
%\section{New update rule}
%\begin{align}
%    & m_{ij}^l=\phi_e\left(h_i^l,h_j^l,\psi(\|x_i^l-x_j^l\|^2), \psi(\langle x_i^l,x_j^l\rangle)\right), \label{eq:mij}
%\end{align}
%\begin{align}
%    & x_i^{l+1}=x_i^l+c\sum_{j \neq i}\phi_x(m^l_{ij})x_j^l \label{eq:x}
%\end{align}
%\begin{align}
    %& m_i^l=\sum_{j\neq i}w_{ij}m^l_{ij} \label{eq:mi}\\
%    &h_i^{l+1}= h_i^{l} + \phi_h(h_i^l, \sum_{j\neq i}m^l_{ij}), \label{eq:h}
%\end{align}

%\begin{align}
    %& m_i^l=\sum_{j\neq i}w_{ij}m^l_{ij} \label{eq:mi}\\
%    &h_i^{l+1}= h_i^{l} + \sum_{j\neq i}\phi_h(m^l_{ij}), \label{eq:h}
%\end{align}

%$(h_i^L,x_i^L)$ as the input of the decoder together.

%To keep $\phi_x$ to be positive, ReLU activation can be considered.

\section{Computational Cost in Comparison to LGN} \label{App:Cost}

\begin{table*}[thp]
\centering
\begin{tabular}{|l|c|c|c|}
\hline
  Model   & \makecell{Time on CPU \\ (ms/batch)} & \makecell{Time on GPU \\ (ms/batch)} & \#Params\\ \hline
LGN     &  $51.4$  & $1.66$ &   4.5k               \\ \hline
LGN (large)    &  $77.0$  & $1.88$ &   12k               \\ \hline
LorentzNet    & $32.9$  & $0.34$ &   224k           \\                     \hline
LorentzNet (small)    & $2.48$  & $0.16$ &   12k       \\                     \hline
\end{tabular}
\caption{Inference time of each model on both CPU and GPU along with their parameter numbers. Models are executed on a cluster with an Intel Xeon CPU E5-2698 v4 and an Nvidia Tesla V100 32GB. Both of the inference times are collected with a batch size of 100.}\label{tab:appcost}
\end{table*}

\begin{table*}[thp]
\centering
\begin{tabular}{|l|c|c|c|c|}
\hline
  Model    & Accuracy & AUC  & \makecell{$1/\varepsilon_B$ \\$(\varepsilon_S=0.5)$} & \makecell{$1/\varepsilon_B$ \\$(\varepsilon_S=0.3)$}\\ \hline
LGN         & $0.929$  & $0.9640$ & $124 \pm 20$      & $435 \pm 95$      \\ \hline
LGN (large)\textsuperscript{\textdagger}        & $-$  & $-$ & $-$      & $-$      \\ \hline
LorentzNet  & $0.942$  & $0.9868$ & ${498 \pm 18}$          & ${2195 \pm 173}$          \\                     \hline
LorentzNet (small)  & ${0.941}$  & ${0.9861}$ & ${428  \pm 29}$          & ${1813 \pm 195}$         \\                     \hline
\end{tabular}
\caption{Performance comparison between LorentzNet and LGN on top tagging dataset. The results for LorentzNet are averaged on 6 runs. The results for LGN are referred to \cite{bogatskiy2020lorentz}. \textsuperscript{\textdagger}We do not evaluate this model since it takes too long to train.}\label{tab:appacc}
\end{table*}

We report the inference time of LorentzNet and LGN with two different scales on both CPU and GPU along with the parameter numbers in Table~\ref{tab:appcost}. The inference time of LorentzNet (small) is $31$ times faster than LGN (large) on CPU and $12$ times faster on GPU, while the number of their trainable parameters is approximately equal. Besides, LorentzNet (small) still reaches $21$ times faster than LGN on CPU and $10$ times faster on GPU, even when the network of LGN is smaller than LorentzNet (small). The comparison under the approximately equal number of parameters shows that LorentzNet is much more efficient.

We also report the performance of LorentzNet and LGN in Table~\ref{tab:appacc}. The LorentzNet (small) still shows a highly competitive performance even when the number of parameters is reduced  to only $5\%$ of the original model. We do not evaluate the performance of LGN (large) due to its slow training process.

\paragraph{Implementation Details.} LorentzNet (small) follows the settings in Sec.~\ref{sec:implement} except all the width of latent space is set to be $16$. LGN (large) follows the settings in \cite{bogatskiy2020lorentz} except the numbers of channels are chosen as $N_{ch}^{(0)} = 6, N_{ch}^{(1)} = 6, N_{ch}^{(2)} = 6, N_{ch}^{(3)} = 6$.
\setcounter{secnumdepth}{2}

\end{document}